# Optimizing Server Load Distribution in Multimedia IoT Environments through LSTM-Based Predictive Algorithms

Somaye Imanpour [a], Ahmadreza Montazerolghaem [b*], Saeed Afshari [c]

[a]: Faculty of Computer Engineering, University of Isfahan, Isfahan, Iran, *s.imanpour@eng.ui.ac.ir*
[b]: Faculty of Computer Engineering, University of Isfahan, Isfahan, Iran, *a.montazerolghaem@comp.ui.ac.ir*
[c]: Faculty of Computer Engineering, Shahreza Campus, University of Isfahan, Isfahan, Iran, *s.afshari@shr.ui.ac.ir*

*A B S T R A C T*

The Internet of Multimedia Things (IoMT) represents a significant advancement in the evolution of IoT technologies, focusing on the transmission and management of multimedia streams. As the volume of data continues to surge and the number of connected devices grows exponentially, internet traffic has reached unprecedented levels, resulting in challenges such as server overloads and deteriorating service quality. Traditional computer network architectures were not designed to accommodate this rapid increase in demand, leading to the necessity for innovative solutions. In response, Software-Defined Networks (SDNs) have emerged as a promising framework, offering enhanced management capabilities by decoupling the control layer from the data layer. This study explores the load balancing of servers within software-defined multimedia IoT networks. The Long Short-Term Memory (LSTM) prediction algorithm is employed to accurately estimate server loads and fuzzy systems are integrated to optimize load distribution across servers. The findings from the simulations indicate that the proposed approach enhances the optimization and management of IoT networks, resulting in improved service quality, reduced operational costs, and increased productivity.

## 1. Introduction

The Internet of Multimedia Things (IoMT) has led to a significant increase in traffic directed toward servers and switches, resulting in congestion and overload within the network. In such conditions, the controller must effectively distribute the load among different servers using load-balancing strategies. Server load balancing requires forecasting methods to enhance load distribution and save energy. Rather than relying on conventional prediction models, this study utilizes the Long Short-Term Memory (LSTM) network, which has demonstrated superior performance in time-series forecasting, particularly in dynamic network environments. Time-series-based forecasting methods are commonly employed for load management in software-defined networks, as these predictions are frequently made for resource management and load balancing. These forecasts rely on historical observations and assist in predicting future values [1,2,3].

The LSTM prediction algorithm, which has been shown to outperform traditional methods such as NLMS and ANN, is utilized in this study to capture complex temporal dependencies. LSTM, a type of Recurrent Neural Network (RNN), is specifically designed to learn long-term dependencies in sequential data. This architecture is particularly effective for processing and predicting complex temporal data, allowing it to identify server load patterns from historical data and provide accurate predictions for future loads. By leveraging LSTM, more precise estimations of server workload are achieved, significantly improving decision-making in dynamic environments[4].

The LSTM prediction algorithm, which has been shown to outperform traditional methods such as NLMS and ANN, is utilized in this study to capture complex temporal dependencies. LSTM, a type of Recurrent Neural Network (RNN), is specifically designed to learn long-term dependencies in sequential data. This architecture is particularly effective for processing and predicting complex temporal data, allowing it to identify server load patterns from historical data and provide accurate predictions for future loads. Compared to other deep learning models such as GRU, CNN-LSTM, and Transformer-based architectures, LSTM offers a balanced trade-off between accuracy and computational efficiency, making it a more practical choice for real-time server load forecasting. While GRU provides a simpler structure, LSTM demonstrates superior performance in capturing long-term dependencies. Transformer-based models, although powerful, require significantly higher computational resources, making them less suitable for real-time SDN environments. Additionally, CNN-LSTM models, which integrate convolutional layers, are more suited for spatial feature extraction rather than sequential network load prediction. Therefore, LSTM is chosen for its robustness in time-series forecasting and its ability to optimize dynamic load balancing strategies effectively. By leveraging LSTM, more precise estimations of server workload are achieved, significantly improving decision-making in dynamic environments [4].

Server load can be determined using threshold values; if a server's load falls below or exceeds a specified amount, it is categorized into a specific group. However, this method lacks the necessary accuracy due to its reliance on a fixed value and disregard for dynamic conditions. Therefore, other methods, such as fuzzy systems, are utilized to adapt to dynamic changes. In contrast to previous studies that implement a three-level fuzzy system [8,13], this study introduces a more adaptive four-level fuzzy system, which enhances load differentiation and optimizes allocation strategies. Fuzzy systems are designed as a decision-making framework based on fuzzy logic to model and manage uncertainty and ambiguity in data. These systems define truth continuously from 0 (non-existence) to 1 (complete existence), allowing for effective modeling of intermediate states [6].

With advancements in technology and the expansion of the multimedia Internet of Things (IoMT), communication between devices and the flow of multimedia data have significantly increased. This surge necessitates optimization in managing server loads. The proposed approach improves load distribution by incorporating four key server resource metrics: CPU, memory, disk, and bandwidth usage. In contrast, most previous studies focused on only two or three parameters, thereby lacking a comprehensive view of server workload. Improper load distribution among servers can lead to decreased efficiency and increased costs. Therefore, this study aims to improve load distribution among servers in software-defined multimedia IoT networks.

The proposed method first calculates each server's resource consumption (CPU, memory, hard disk, and bandwidth). Then, resource consumption forecasting for the server is conducted using artificial intelligence algorithms. By utilizing LSTM's predictive capabilities, the system proactively manages traffic allocation before congestion occurs, significantly enhancing network efficiency. Based on the obtained predictions, the server is categorized at a specific level, indicating its position within multiple threshold levels. Servers with lower levels are prioritized for receiving load. Subsequently, it is assessed whether a server is in an overloaded state. If an overload is present, the load is transferred from the overloaded server to a server with a lower load. The main contributions of this paper are as follows:
1. Utilizing four metrics to calculate server load, including CPU usage, memory, disk, and bandwidth.
2. Employing Long Short-Term Memory (LSTM) algorithms for predicting server loads.
3. Using fuzzy systems for accurate categorization of servers into four levels.

The subsequent sections of the paper will include a review of related work in Section 2, presentation of the proposed system in Section 3, simulation results in Section 4, and concluding remarks in Section 5.

## 2. Related works

Dynamic algorithms for balancing server loads initially assess the current conditions of the servers before determining which server should handle incoming traffic based on its state. Zhang et al. [7] introduced a dynamic load-balancing strategy that relies on the response time of servers, termed Load Balancing Based on Server Response Time (LBBSRT). This strategy allocates loads based on response times, prioritizing servers that exhibit quicker or more consistent response times. Compared to traditional approaches, this technique enhances overall performance and is more cost-effective, as it requires less hardware and allows for software-based adaptation.

Montazerolghaem [8] developed a framework to address resource management challenges in multimedia Internet of Things (IoMT) networks. This framework employs function virtualization technology for all servers, enabling the controller to adjust the network's scale. By applying the normalized minimum mean square prediction method, which analyzes historical resource utilization data (including CPU, memory, and disk usage), servers are classified into three categories through predictive analysis and fuzzy logic. This classification process assists in evaluating the overall network condition and facilitates appropriate adjustments to network size.

The research presented in [9] focuses on load balancing within software-defined networks by utilizing machine learning methods such as artificial neural networks (ANN) and reinforcement learning. These techniques facilitate server load balancing and network traffic routing. The artificial neural network learns how to allocate traffic to servers by leveraging insights gained from past interactions.

Article [10] introduces a load-balancing algorithm specifically designed for data center networks to enhance server efficiency and performance. By leveraging software-defined networking (SDN) and artificial intelligence methods, the study seeks to manage high traffic volumes effectively, mitigate data loss, and ensure continuous server availability.

In [11], a PID controller integrated with neural networks is proposed to enhance server load balancing in SDN environments. This PID controller fine-tunes the polling frequency to assess server response times, relying on two threshold parameters: a low-load threshold and a high-load threshold. These thresholds determine the server's load status based on variations in response times, categorizing it into low, medium, or high-load classes. This adaptive classification allows for optimal server selection based on response times and load, thereby improving the load-balancing process.

The study presented in [12] explores a hybrid approach for balancing loads between servers and connections within distributed storage systems via software-defined networks (SDN). The load-balancing algorithm ensures that traffic distribution aligns with overall network conditions and storage server loads.

In [13], the challenges associated with efficient resource allocation for multimedia streaming in Software-Defined Internet of Vehicles (IoV) networks are addressed. Load balancing in IoV networks is achieved through the implementation of the ELQ2 framework. By employing intelligent algorithms and a modular architecture, ELQ2 dynamically selects physical machines (PMs) with the least load for multimedia content delivery, thereby optimizing resource distribution across the network.

Article [14] focuses on optimizing software-defined multimedia frameworks to enhance networking capabilities in multimedia-over-IP (MoIP) environments. The research outlines various load-balancing strategies employed to manage traffic across servers within MoIP networks. Three specific algorithms for load distribution are highlighted: call-join-shortest-queue

(CJSQ), transaction-join-shortest-queue (TJSQ), and transaction-least-work-left (TLWL). These algorithms aim to distribute workloads efficiently among servers by directing new calls or transactions based on the servers' current load conditions.

The study in [15] presents a framework called GreenVoIP, which effectively addresses the issue of load balancing among servers in VoIP networks. By utilizing cloud computing principles alongside SDN and NFV technologies, GreenVoIP enables dynamic resource management that responds to fluctuating demand in real time. This framework allows for the efficient allocation of VoIP servers based on current network loads, thereby preventing server overload during peak times and minimizing energy consumption during periods of lower demand. Specifically, when high traffic is detected, additional virtualized server resources can be activated to handle the load, ensuring that QoS requirements are met without compromising service quality. Conversely, GreenVoIP can deactivate or allocate less active resources during periods of reduced demand, thereby conserving energy. This resource management adaptability enhances operational efficiency and contributes to the goal of green computing by reducing the overall energy footprint of VoIP services.

The study in [16] discusses the benefits and challenges of implementing Software-Defined Networking (SDN), particularly in hybrid environments where both SDN-enabled and traditional devices coexist. Due to budget constraints and technical challenges, many network providers opt for incremental SDN deployment rather than full implementation. This paper introduces a new load-balancing scheme called LBORU (Load Balancing by Optimizing Resource Utilization), designed specifically for hybrid SDN networks. This approach utilizes a minimal set of SDN components, including a controller and a switch. The proposed load-balancing scheme continuously monitors server load indicators—such as CPU load, I/O read, I/O write, link upload, and link download—and employs multi-parameter metrics to schedule connections effectively. This strategy enables more efficient load distribution across servers, addressing resource overload issues and ensuring better Quality of Experience (QoE) for users. The results demonstrate that LBORU outperforms traditional load-balancing methods, including Random, Round-Robin, and Weighted Fair Queuing, as well as existing server response time-based schemes.

The integration of edge and fog computing in healthcare systems offers real-time processing capabilities, yet load balancing remains a critical challenge. Efficient workload distribution across edge servers is essential to prevent bottlenecks, reduce latency, and optimize resource utilization. Load-balancing techniques in SDN-based healthcare networks leverage predictive analytics and dynamic traffic management to enhance system performance. The combination of static and dynamic load-balancing strategies, including SDN-driven approaches, allows adaptive workload distribution that responds to varying network conditions. Implementing an effective load-balancing framework in edge/fog-based healthcare environments ensures improved QoS, reduced system latency, and cost-efficient resource management, ultimately leading to better patient care and optimized healthcare service delivery [17].

Load balancing in Data Center Networks (DCN) is essential for optimizing server performance, minimizing latency, and ensuring efficient resource utilization, particularly as internet applications grow. Software-Defined Networking (SDN) enables dynamic load balancing by centralizing control and improving traffic management. Various techniques, including threshold-based, heuristic, stochastic load prediction, and hybrid machine learning algorithms like Multiple Regression-Based Searching (MRBS), are employed to distribute network traffic efficiently. SDN-based approaches, such as controller-driven load management and Equal-Cost Multipath (ECMP) routing, improve scalability but face challenges like real-time decision-making, controller bottlenecks, and handling heterogeneous workloads[18].

Table I provides a summary of the different references, detailing their methods for server load balancing, the metrics utilized to evaluate server load, the prediction algorithms applied, and the types of thresholds implemented.

Table 1. A Review of Article

| Reference | Server Load Measurement Metric | Prediction Algorithm | Threshold Type | Energy Consumption |
|---|---|---|---|---|
| [7] | Response Time | | Single Threshold - Two-Level | - |
| [8] | CPU, Memory, Disk Usage | NLMS | Fuzzy System - Three-Level | ✓ |
| [9] | CPU, Memory Usage | - | - | - |
| [10] | CPU, Memory Usage, Number of Requests Processed per Second | - | Single Threshold - Two-Level | - |
| [11] | Response Time | - | Two Thresholds | - |

| | | | - Three-Level | |
|---|---|---|---|---|
| [12] | Number of Requests Processed per Server | - | - | - |
| [13] | CPU, Memory, Disk Usage | NLMS | Fuzzy System - Three-Level | ✓ |
| [14] | CPU, Memory Usage | - | - | - |
| [15] | CPU, Memory Usage | NLMS | Fuzzy System - Three-Level | ✓ |
| [16] | CPU, Bandwidth Usage | - | Single Threshold - Two-Level | - |
| [17] | Response Time | - | Single Threshold - Two-Level | - |
| [18] | Response Time, Bandwidth | Multiple Regression | Multi Threshold | - |
| Proposed Approach | CPU, Memory, Disk, and Bandwidth Usage | LSTM | Fuzzy System - Four-Level | ✓ |

## 3. Proposed framework

As shown in Figure 1, the software-defined network divides the Internet of Things (IoT) network into three layers. The first layer is the application layer, which includes applications that are added to the controller through these applications. The second layer is the control layer, which comprises the controller, while the third layer is the data layer, which includes servers, switches, and links tables and Figures.

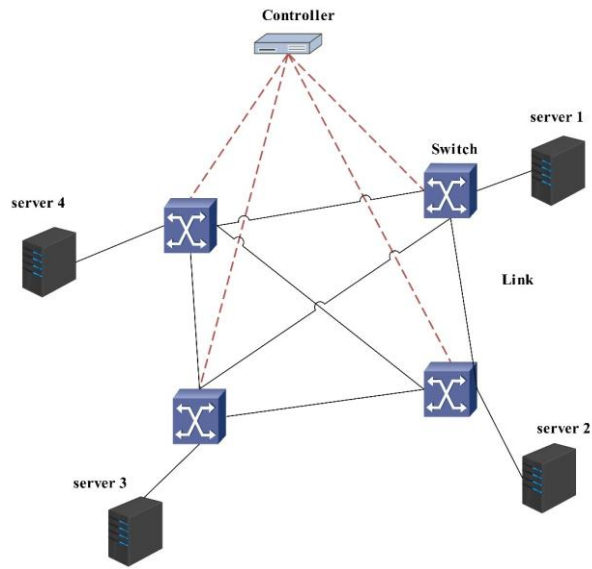

Figure. 1. Software-Defined Multimedia Internet of things Network Model

As illustrated in Figure 2, the proposed method algorithm consists of three main components. The first component is the load balancing of servers in the software-defined IoMT network. To achieve load balancing among the servers, the server load is first measured based on criteria including CPU usage, memory, hard disk, and bandwidth consumption. Then, using a long-term short-term memory prediction algorithm, the server loads are forecasted. With the help of a fuzzy system, the servers are classified, and load-balancing operations are carried out based on the server's classification level.

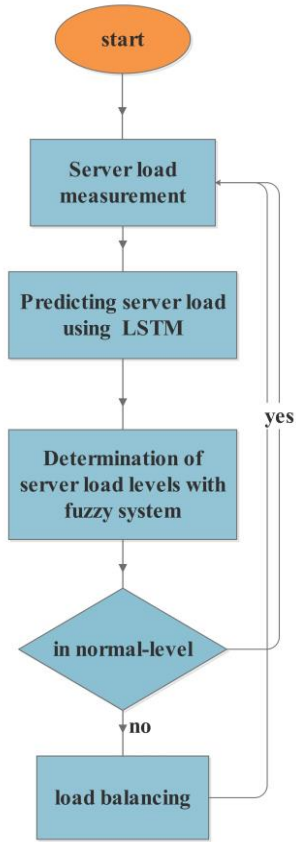

Figure. 2. Flowchart of the Proposed Method Algorithm

### 3.1. System Model

The network comprises switches, servers, controllers, and links. Let $H=\{h_1, h_2, ..., h_i, ..., h_M\}$ represent the set of servers, where $M$ indicates the total number of servers in the network, and $h_i$ denotes the ith server. For clarity, Table 2 provides the definitions and notations used in this study.

Table 2. The Summary of the Notations that are used.

| Symbols | Description |
|---|---|
| S | Set of switches |
| N | Total number of switches |
| H | Set of servers |
| M | Total number of servers |
| $x_i$ | CPU consumption of server i |
| $y_i$ | Memory consumption of server i |
| $z_i$ | Hard disk consumption of server i |
| $w_i$ | Bandwidth consumption of server i |
| $Ld_{CPU}$ | CPU consumption of server i at time t |
| $Ld_{mem}$ | Memory consumption of server i at time t |
| $Ld_{hdd}$ | Hard disk consumption of server i at time t |
| $Ld_{bw}$ | Bandwidth consumption of server i at time t |
| $\vartheta_i^x$ | CPU resource capacity of server i |
| $\vartheta_i^y$ | Memory resource capacity of server i |
| $\vartheta_i^z$ | Bandwidth resource capacity of server i |
| $\vartheta_i^w$ | Hard disk resource capacity of server i |
| $f_{CPU}$ | Matrix of CPU consumption by servers |
| $f_{mem}$ | Matrix of memory consumption by servers |

| $f_{hdd}$ | Matrix of bandwidth consumption by servers |
| $f_{bw}$ | Matrix of hard disk consumption by servers |

*Server Load Measurement*

In the first step of server load balancing, the controller measures the load of the servers within its domain. The controller evaluates the server load based on CPU, memory, bandwidth, and hard disk consumption. To measure the server load, the current CPU usage of the server is divided by the CPU resource capacity, resulting in a value between zero and one, represented as $x_i$ as show in Equation (1). Similarly, for calculating the memory, hard disk, and bandwidth consumption of the server, the same approach is applied, with the results denoted as $y_i$, $z_i$, and $w_i$, as show in Equation (2), (3), and (4), respectively. This method ensures a normalized representation of resource utilization, facilitating effective load-balancing decisions.

$$x_i = \frac{Ld_{CPU}}{\vartheta_i^x} \qquad 0 \leq x_i \leq 1 \qquad (1)$$

$$y_i = \frac{Ld_{mem}}{\vartheta_i^y} \qquad 0 \leq y_i \leq 1 \qquad (2)$$

$$z_i = \frac{Ld_{hdd}}{\vartheta_i^z} \qquad 0 \leq z_i \leq 1 \qquad (3)$$

$$w_i = \frac{Ld_{bw}}{\vartheta_i^w} \qquad 0 \leq w_i \leq 1 \qquad (4)$$

To predict the server load in the next step, the long short-term memory (LSTM) prediction algorithm is utilized. This requires the calculation of a matrix representing the resource consumption of all servers within the domain. The metrics for measuring server load include CPU usage, memory usage, bandwidth, and hard disk consumption. Therefore, the prediction of each of these metrics is calculated for all servers. The matrix $f_{h_i}$, as shown in Formula (5), represents the resource consumption of the servers. The size of this matrix is $P*W$, where $P = 4$. Each row of the matrix corresponds to one of the consumption metrics: CPU, memory, hard disk, and bandwidth, while $W$ denotes the size of the window.

This structured approach allows for a systematic analysis and forecasting of the resource utilization patterns of the servers, facilitating better load-balancing decisions in the network.

$$f_{h_i} = \begin{bmatrix} x_{11} & x_{12} & \ldots & x_{1W} \\ \ldots & \ldots & \ldots & \ldots \\ x_{P1} & x_{P1} & \ldots & x_{PW} \end{bmatrix} \qquad (5)$$

$$if\ \acute{M} < P\ then\ x_{PW} = 0\ , p = 4\ , w = 1,2,\ldots,W$$

*Server Load Prediction Using Long Short-Term Memory (LSTM)*

In the load balancing of software-defined networks, forecasting methods are employed to predict controller load, server load, and traffic. This ultimately leads to load balancing and energy savings. In this section, prediction techniques are utilized specifically for forecasting server loads. Various methods exist for prediction, but time-series-based forecasting methods are more commonly used in software-defined networks. This is because time-series forecasting samples a specific metric periodically at fixed intervals, resulting in a time series that includes a sequence of recent observations. Time-series methods utilize this sequence to predict future values.

Long Short-Term Memory (LSTM), a specific architecture of recurrent neural networks (RNN), is employed to predict server loads. LSTM is suitable for learning from experience for classification, processing, and predicting time series with unknown time lags.

- Model Architecture Details

The LSTM model is implemented using a sequential structure, consisting of a single LSTM layer with five neurons and a Dense output layer with one neuron. The number of lookback steps varies depending on the predicted metric. For CPU usage, disk usage, and bandwidth prediction, a lookback of five is used, while for memory prediction, a lookback of two is applied. These values were determined through a trial-and-error approach to find the most suitable configuration. The model is trained using the Adam optimizer, with Mean Squared Error (MSE) as the loss function. Training is performed with 15 epochs and a batch size of one.

The input data is first normalized and then divided into two sets: a training set 80% and a testing set 20%. The model is trained using time-series data with a lookback window of size $n$.

The model operates in two phases: training and forecasting. In the training phase, the training data is fed into the model, and the parameters of the neural network are dynamically adjusted to achieve the desired output. This process is carried out through the backpropagation algorithm, which propagates the computed error from the output to the input and updates the model weights to minimize the error to the lowest possible value.

In the prediction phase, the model is tested with new, previously unseen data. The expected output is to predict the resource consumption levels for all servers in the domain, represented as $\hat{x}_{t+1}^i$, $\hat{y}_{t+1}^i$, $\hat{z}_{t+1}^i$, and $\hat{w}_{t+1}^i$. The LSTM model, with its ability to learn complex temporal dependencies, provides a powerful tool for predicting load in software-defined networks. This model not only aids in improving load balancing but also plays a crucial role in optimizing resource and energy consumption. Utilizing LSTM for server load prediction enables more precise resource management and enhances the efficiency of software-defined systems.

### *Determining Server Load Levels Using Fuzzy Systems*

A fuzzy system is a mathematical model based on the theory of fuzzy sets, used for modeling and analyzing issues that involve a combination of uncertainty and continuity. Fuzzy systems can be highly effective in situations where the relationships among variables require abstraction and innovation. Fuzzy algorithms are inspired by human abstract reasoning and can effectively model information that possesses uncertainty and complexity.

Server classification using a fuzzy system depends on the predicted resource consumption of the server, represented as ($\hat{x}_{t+1}^i$, $\hat{y}_{t+1}^i$, $\hat{z}_{t+1}^i$, $\hat{w}_{t+1}^i$). Servers are categorized into four levels: under-load, normal-load, highly-load, and over-load. The input for this classification consists of ($\hat{x}_{t+1}^i$, $\hat{y}_{t+1}^i$, $\hat{z}_{t+1}^i$, $\hat{w}_{t+1}^i$), and the output will be one of the four mentioned levels. Each step of the process is explained in detail below:

- Variable Definition:

At this stage, various features that may be used for server classification are identified. In this study, the predicted resource consumption of servers is utilized, including CPU usage, memory, hard disk, and bandwidth.

- Membership Functions Definition:

As shown in Figure 3-6, a fuzzy membership function is defined for each feature. These functions assist in transforming numerical features into fuzzy sets (low, medium, high).

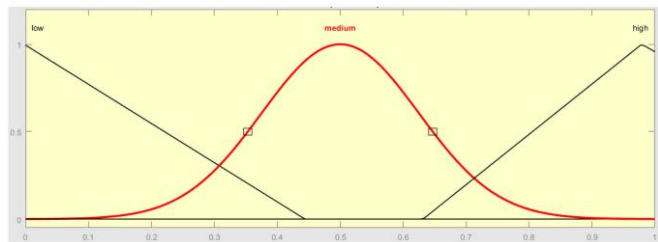

Figure. 3.    Membership Function Input for $\hat{x}_{t+1}^i$ in the Range [0,1]}

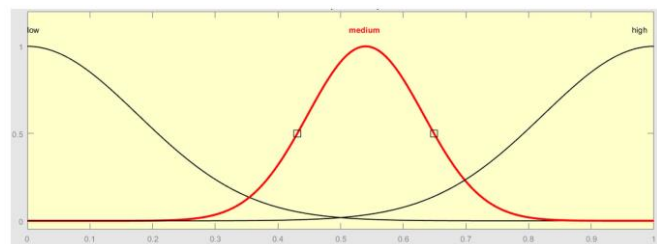

Figure. 4.    Membership Function Input for $\hat{y}_{t+1}^i$ in the Range [0,1]}

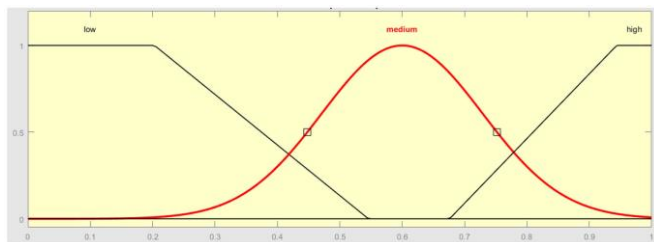

Figure. 5.    Membership Function Input for $\hat{z}_{t+1}^i$ in the Range [0,1]}

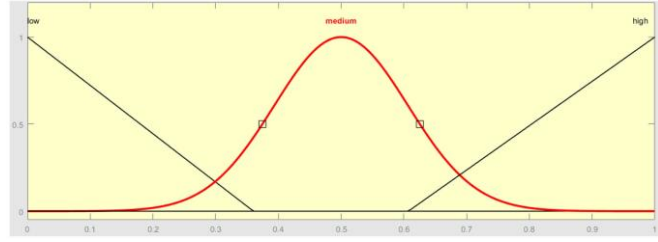

Figure. 6.  Membership Function Input for $\hat{w}_{t+1}^i$ in the Range [0,1]}

- Fuzzy Rules Definition:

The fuzzy rules necessary for classifying servers are defined. These rules encompass conditions that depend on the features and their combinations. The system comprises 81 rules defined in an "if-then" format. These rules represent combinations of input states that specify the corresponding outputs. For example, if all inputs are in the "low" state, the output will also be in the "under-load" state. This fuzzy classification system allows for a nuanced understanding of server loads, enabling more effective load-balancing strategies within the network.

1) When ($\hat{x}_{t+1}^i$ is classified as low ) and ($\hat{y}_{t+1}^i$ is classified as low ) and ($\hat{z}_{t+1}^i$ is classified as low ) and ($\hat{w}_{t+1}^i$ is classified as low ), then (classk is categorized as Under-loaded).
2) When ($\hat{x}_{t+1}^i$ is classified as low ) and ($\hat{y}_{t+1}^i$ is classified as low ) and ($\hat{z}_{t+1}^i$ is classified as low) and ($\hat{w}_{t+1}^i$ is classified as Medium ), then (classk is categorized as Under -loaded).
3) When ($\hat{x}_{t+1}^i$ is classified as low ) and ($\hat{y}_{t+1}^i$ is classified as low ) and ($\hat{z}_{t+1}^i$ is classified as Medium) and ($\hat{w}_{t+1}^i$ is classified as low ), then (classk is categorized as Under -loaded).
4) When ($\hat{x}_{t+1}^i$ is classified as low ) and ($\hat{y}_{t+1}^i$ is classified as Medium ) and ($\hat{z}_{t+1}^i$ is classified as low ) and ($\hat{w}_{t+1}^i$ is classified as low ), then (classk is categorized as Under -loaded).
5) When ($\hat{x}_{t+1}^i$ is classified as Medium ) and ($\hat{y}_{t+1}^i$ is classified as low) and ($\hat{z}_{t+1}^i$ is classified as low ) and ($\hat{w}_{t+1}^i$ is classified as  low), then (classk is categorized as Under -loaded).
6) When ($\hat{x}_{t+1}^i$ is classified as low) and ($\hat{y}_{t+1}^i$ is classified as low ) and ($\hat{z}_{t+1}^i$ is classified as Medium) and ($\hat{w}_{t+1}^i$ is classified as Medium ), then (classk is categorized as Under -loaded).
7) When ($\hat{x}_{t+1}^i$ is classified as low ) and ($\hat{y}_{t+1}^i$ is classified as Medium) and ($\hat{z}_{t+1}^i$ is classified as Medium ) and ($\hat{w}_{t+1}^i$ is classified as  low), then (classk is categorized as Under -loaded).
8) When ($\hat{x}_{t+1}^i$ is classified as Medium ) and ($\hat{y}_{t+1}^i$ is classified as Medium ) and ($\hat{z}_{t+1}^i$ is classified as low ) and ($\hat{w}_{t+1}^i$ is classified as low ), then (classk is categorized as Under -loaded).
9) When ($\hat{x}_{t+1}^i$ is classified as low ) and ($\hat{y}_{t+1}^i$ is classified as Medium ) and ($\hat{z}_{t+1}^i$ is classified as low ) and ($\hat{w}_{t+1}^i$ is classified as Medium ), then (classk is categorized as Under -loaded).
10) When ($\hat{x}_{t+1}^i$ is classified as Medium) and ($\hat{y}_{t+1}^i$ is classified as low ) and ($\hat{z}_{t+1}^i$ is classified as Medium ) and ($\hat{w}_{t+1}^i$ is classified as low ), then (classk is categorized as Under -loaded).
11) When ($\hat{x}_{t+1}^i$ is classified as Medium ) and ($\hat{y}_{t+1}^i$ is classified as low ) and ($\hat{z}_{t+1}^i$ is classified as low ) and ($\hat{w}_{t+1}^i$ is classified as Medium ), then (classk is categorized as Under -loaded).
12) When ($\hat{x}_{t+1}^i$ is classified as low ) and ($\hat{y}_{t+1}^i$ is classified as low) and ($\hat{z}_{t+1}^i$ is classified as low ) and ($\hat{w}_{t+1}^i$ is classified as High ), then (classk is categorized as Normal -loaded).
13) When ($\hat{x}_{t+1}^i$ is classified as low ) and ($\hat{y}_{t+1}^i$ is classified as low) and ($\hat{z}_{t+1}^i$ is classified as High ) and ($\hat{w}_{t+1}^i$ is classified as low ), then (classk is categorized as Normal -loaded).
14) When ($\hat{x}_{t+1}^i$ is classified as low ) and ($\hat{y}_{t+1}^i$ is classified as High ) and ($\hat{z}_{t+1}^i$ is classified as low ) and ($\hat{w}_{t+1}^i$ is classified as low ), then (classk is categorized as Normal -loaded).
15) When ($\hat{x}_{t+1}^i$ is classified as High) and ($\hat{y}_{t+1}^i$ is classified as low ) and ($\hat{z}_{t+1}^i$ is classified as low ) and ($\hat{w}_{t+1}^i$ is classified as low ), then (classk is categorized as Normal -loaded).
16) When ($\hat{x}_{t+1}^i$ is classified as low ) and ($\hat{y}_{t+1}^i$ is classified as  Medium) and ($\hat{z}_{t+1}^i$ is classified as Medium  ) and ($\hat{w}_{t+1}^i$ is classified as Medium ), then (classk is categorized as Normal -loaded).
17) When ($\hat{x}_{t+1}^i$ is classified as Medium ) and ($\hat{y}_{t+1}^i$ is classified as low ) and ($\hat{z}_{t+1}^i$ is classified as Medium) and ($\hat{w}_{t+1}^i$ is classified as Medium ), then (classk is categorized as Normal -loaded).
18) When ($\hat{x}_{t+1}^i$ is classified as  Medium) and ($\hat{y}_{t+1}^i$ is classified as Medium ) and ($\hat{z}_{t+1}^i$ is classified as low ) and ($\hat{w}_{t+1}^i$ is classified as Medium ), then (classk is categorized as Normal -loaded).
19) When ($\hat{x}_{t+1}^i$ is classified as Medium ) and ($\hat{y}_{t+1}^i$ is classified as Medium ) and ($\hat{z}_{t+1}^i$ is classified as Medium ) and ($\hat{w}_{t+1}^i$ is classified as low ), then (classk is categorized as Normal -loaded).
20) When ($\hat{x}_{t+1}^i$ is classified as Medium ) and ($\hat{y}_{t+1}^i$ is classified as Medium ) and ($\hat{z}_{t+1}^i$ is classified as Medium) and ($\hat{w}_{t+1}^i$ is classified as Medium ), then (classk is categorized as Normal -loaded).
21) When ($\hat{x}_{t+1}^i$ is classified as High ) and ($\hat{y}_{t+1}^i$ is classified as low ) and ($\hat{z}_{t+1}^i$ is classified as Medium  ) and ($\hat{w}_{t+1}^i$ is classified as low ), then (classk is categorized as  Normal -loaded).
22) When ($\hat{x}_{t+1}^i$ is classified as  High ) and ($\hat{y}_{t+1}^i$ is classified as low ) and ($\hat{z}_{t+1}^i$ is classified as low ) and ($\hat{w}_{t+1}^i$ is classified as Medium), then (classk is categorized as Normal -loaded).
23) When ($\hat{x}_{t+1}^i$ is classified as Medium) and ($\hat{y}_{t+1}^i$ is classified as High ) and ($\hat{z}_{t+1}^i$ is classified as low ) and ($\hat{w}_{t+1}^i$ is classified as low ), then (classk is categorized as Normal -loaded).
24) When ($\hat{x}_{t+1}^i$ is classified as Medium) and ($\hat{y}_{t+1}^i$ is classified as low ) and ($\hat{z}_{t+1}^i$ is classified as High ) and ($\hat{w}_{t+1}^i$ is classified as low ), then (classk is categorized as Normal -loaded).
25) When ($\hat{x}_{t+1}^i$ is classified as Medium ) and ($\hat{y}_{t+1}^i$ is classified as low ) and ($\hat{z}_{t+1}^i$ is classified as low ) and ($\hat{w}_{t+1}^i$ is classified as High ), then (classk is categorized as Normal -loaded).
26) When ($\hat{x}_{t+1}^i$ is classified as low ) and ($\hat{y}_{t+1}^i$ is classified as High ) and ($\hat{z}_{t+1}^i$ is classified as low ) and ($\hat{w}_{t+1}^i$ is classified as Medium ), then (classk is categorized as Normal -loaded).

27) When ($\hat{x}_{t+1}^i$ is classified as low) and ($\hat{y}_{t+1}^i$ is classified as High) and ($\hat{z}_{t+1}^i$ is classified as Medium ) and ($\hat{w}_{t+1}^i$ is classified as low ), then (classk is categorized as Normal -loaded).
28) When ($\hat{x}_{t+1}^i$ is classified as low ) and ($\hat{y}_{t+1}^i$ is classified as Medium) and ($\hat{z}_{t+1}^i$ is classified as low  ) and ($\hat{w}_{t+1}^i$ is classified as High ), then (classk is categorized as Normal -loaded).
29) When ($\hat{x}_{t+1}^i$ is classified as low ) and ($\hat{y}_{t+1}^i$ is classified as Medium ) and ($\hat{z}_{t+1}^i$ is classified as High) and ($\hat{w}_{t+1}^i$ is classified as  low), then (classk is categorized as Normal -loaded).
30) When ($\hat{x}_{t+1}^i$ is classified as low ) and ($\hat{y}_{t+1}^i$ is classified as low ) and ($\hat{z}_{t+1}^i$ is classified as Medium ) and ($\hat{w}_{t+1}^i$ is classified as High ), then (classk is categorized as Normal -loaded).
31) When ($\hat{x}_{t+1}^i$ is classified as low ) and ($\hat{y}_{t+1}^i$ is classified as low ) and ($\hat{z}_{t+1}^i$ is classified as High ) and ($\hat{w}_{t+1}^i$ is classified as Medium ), then (classk is categorized as Normal -loaded).
32) When ($\hat{x}_{t+1}^i$ is classified as High ) and ($\hat{y}_{t+1}^i$ is classified as Medium) and ($\hat{z}_{t+1}^i$ is classified as low ) and ($\hat{w}_{t+1}^i$ is classified as low ), then (classk is categorized as Normal -loaded).
33) When ($\hat{x}_{t+1}^i$ is classified as low) and ($\hat{y}_{t+1}^i$ is classified as High ) and ($\hat{z}_{t+1}^i$ is classified as Medium ) and ($\hat{w}_{t+1}^i$ is classified as Medium ), then (classk is categorized as Normal -loaded).
34) When ($\hat{x}_{t+1}^i$ is classified as low ) and ($\hat{y}_{t+1}^i$ is classified as Medium) and ($\hat{z}_{t+1}^i$ is classified as High ) and ($\hat{w}_{t+1}^i$ is classified as Medium ), then (classk is categorized as Normal -loaded).
35) When ($\hat{x}_{t+1}^i$ is classified as low ) and ($\hat{y}_{t+1}^i$ is classified as Medium) and ($\hat{z}_{t+1}^i$ is classified as  Medium) and ($\hat{w}_{t+1}^i$ is classified as High ), then (classk is categorized as Normal -loaded).
36) When ($\hat{x}_{t+1}^i$ is classified as High ) and ($\hat{y}_{t+1}^i$ is classified as low ) and ($\hat{z}_{t+1}^i$ is classified as Medium  ) and ($\hat{w}_{t+1}^i$ is classified as Medium ), then (classk is categorized as  Normal -loaded).
37) When ($\hat{x}_{t+1}^i$ is classified as High ) and ($\hat{y}_{t+1}^i$ is classified as Medium) and ($\hat{z}_{t+1}^i$ is classified as  low ) and ($\hat{w}_{t+1}^i$ is classified as Medium ), then (classk is categorized as Normal -loaded).
38) When ($\hat{x}_{t+1}^i$ is classified as High ) and ($\hat{y}_{t+1}^i$ is classified as Medium ) and ($\hat{z}_{t+1}^i$ is classified as Medium ) and ($\hat{w}_{t+1}^i$ is classified as low ), then (classk is categorized as Normal -loaded).
39) When ($\hat{x}_{t+1}^i$ is classified as Medium ) and ($\hat{y}_{t+1}^i$ is classified as low ) and ($\hat{z}_{t+1}^i$ is classified as  Medium) and ($\hat{w}_{t+1}^i$ is classified as High ), then (classk is categorized as Normal -loaded).
40) When ($\hat{x}_{t+1}^i$ is classified as Medium ) and ($\hat{y}_{t+1}^i$ is classified as low) and ($\hat{z}_{t+1}^i$ is classified as High ) and ($\hat{w}_{t+1}^i$ is classified as Medium), then (classk is categorized as Normal -loaded).
41) When ($\hat{x}_{t+1}^i$ is classified as Medium ) and ($\hat{y}_{t+1}^i$ is classified as High ) and ($\hat{z}_{t+1}^i$ is classified as Medium  ) and ($\hat{w}_{t+1}^i$ is classified as low ), then (classk is categorized as  Normal -loaded).
42) When ($\hat{x}_{t+1}^i$ is classified as Medium ) and ($\hat{y}_{t+1}^i$ is classified as High ) and ($\hat{z}_{t+1}^i$ is classified as low ) and ($\hat{w}_{t+1}^i$ is classified as Medium), then (classk is categorized as Normal -loaded).
43) When ($\hat{x}_{t+1}^i$ is classified as Medium ) and ($\hat{y}_{t+1}^i$ is classified as Medium ) and ($\hat{z}_{t+1}^i$ is classified as High ) and ($\hat{w}_{t+1}^i$ is classified as low ), then (classk is categorized as Normal -loaded).
44) When ($\hat{x}_{t+1}^i$ is classified as Medium ) and ($\hat{y}_{t+1}^i$ is classified as Medium) and ($\hat{z}_{t+1}^i$ is classified as low ) and ($\hat{w}_{t+1}^i$ is classified as High ), then (classk is categorized as Normal -loaded).
45) When ($\hat{x}_{t+1}^i$ is classified as low ) and ($\hat{y}_{t+1}^i$ is classified as low ) and ($\hat{z}_{t+1}^i$ is classified as High ) and ($\hat{w}_{t+1}^i$ is classified as High ), then (classk is categorized as High -loaded).
46) When ($\hat{x}_{t+1}^i$ is classified as High ) and ($\hat{y}_{t+1}^i$ is classified as High ) and ($\hat{z}_{t+1}^i$ is classified as low ) and ($\hat{w}_{t+1}^i$ is classified as low ), then (classk is categorized as High -loaded).
47) When ($\hat{x}_{t+1}^i$ is classified as low ) and ($\hat{y}_{t+1}^i$ is classified as High ) and ($\hat{z}_{t+1}^i$ is classified as low ) and ($\hat{w}_{t+1}^i$ is classified as High ), then (classk is categorized as High -loaded).
48) When ($\hat{x}_{t+1}^i$ is classified as High ) and ($\hat{y}_{t+1}^i$ is classified as low ) and ($\hat{z}_{t+1}^i$ is classified as High  ) and ($\hat{w}_{t+1}^i$ is classified as low), then (classk is categorized as  High -loaded).
49) When ($\hat{x}_{t+1}^i$ is classified as High) and ($\hat{y}_{t+1}^i$ is classified as low ) and ($\hat{z}_{t+1}^i$ is classified as low ) and ($\hat{w}_{t+1}^i$ is classified as High ), then (classk is categorized as  High -loaded).
50) When ($\hat{x}_{t+1}^i$ is classified as low ) and ($\hat{y}_{t+1}^i$ is classified as High) and ($\hat{z}_{t+1}^i$ is classified as High ) and ($\hat{w}_{t+1}^i$ is classified as low ), then (classk is categorized as  High -loaded).
51) When ($\hat{x}_{t+1}^i$ is classified as Medium ) and ($\hat{y}_{t+1}^i$ is classified as Medium ) and ($\hat{z}_{t+1}^i$ is classified as Medium) and ($\hat{w}_{t+1}^i$ is classified as High ), then (classk is categorized as  High -loaded).
52) When ($\hat{x}_{t+1}^i$ is classified as Medium ) and ($\hat{y}_{t+1}^i$ is classified as Medium ) and ($\hat{z}_{t+1}^i$ is classified as High ) and ($\hat{w}_{t+1}^i$ is classified as Medium ), then (classk is categorized as High -loaded).
53) When ($\hat{x}_{t+1}^i$ is classified as Medium) and ($\hat{y}_{t+1}^i$ is classified as High ) and ($\hat{z}_{t+1}^i$ is classified as Medium ) and ($\hat{w}_{t+1}^i$ is classified as Medium ), then (classk is categorized as High -loaded).
54) When ($\hat{x}_{t+1}^i$ is classified as High ) and ($\hat{y}_{t+1}^i$ is classified as Medium ) and ($\hat{z}_{t+1}^i$ is classified as Medium ) and ($\hat{w}_{t+1}^i$ is classified as Medium ), then (classk is categorized as High -loaded).
55) When ($\hat{x}_{t+1}^i$ is classified as low ) and ($\hat{y}_{t+1}^i$ is classified as High) and ($\hat{z}_{t+1}^i$ is classified as Medium ) and ($\hat{w}_{t+1}^i$ is classified as High ), then (classk is categorized as High -loaded).
56) When ($\hat{x}_{t+1}^i$ is classified as Medium) and ($\hat{y}_{t+1}^i$ is classified as low ) and ($\hat{z}_{t+1}^i$ is classified as High  ) and ($\hat{w}_{t+1}^i$ is classified as High ), then (classk is categorized as  High -loaded).
57) When ($\hat{x}_{t+1}^i$ is classified as low ) and ($\hat{y}_{t+1}^i$ is classified as High ) and ($\hat{z}_{t+1}^i$ is classified as Medium ) and ($\hat{w}_{t+1}^i$ is classified as High ), then (classk is categorized as High -loaded).
58) When ($\hat{x}_{t+1}^i$ is classified as High ) and ($\hat{y}_{t+1}^i$ is classified as low ) and ($\hat{z}_{t+1}^i$ is classified as Medium ) and ($\hat{w}_{t+1}^i$ is classified as High ), then (classk is categorized as High -loaded).
59) When ($\hat{x}_{t+1}^i$ is classified as High ) and ($\hat{y}_{t+1}^i$ is classified as low) and ($\hat{z}_{t+1}^i$ is classified as High ) and ($\hat{w}_{t+1}^i$ is classified as Medium ), then (classk is categorized as High -loaded).
60) When ($\hat{x}_{t+1}^i$ is classified as low ) and ($\hat{y}_{t+1}^i$ is classified as High ) and ($\hat{z}_{t+1}^i$ is classified as High ) and ($\hat{w}_{t+1}^i$ is classified as Medium ), then (classk is categorized as High -loaded).
61) When ($\hat{x}_{t+1}^i$ is classified as Medium ) and ($\hat{y}_{t+1}^i$ is classified as High ) and ($\hat{z}_{t+1}^i$ is classified as low ) and ($\hat{w}_{t+1}^i$ is classified as High), then (classk is categorized as High -loaded).
62) When ($\hat{x}_{t+1}^i$ is classified as Medium ) and ($\hat{y}_{t+1}^i$ is classified as High ) and ($\hat{z}_{t+1}^i$ is classified as High ) and ($\hat{w}_{t+1}^i$ is classified as low ), then (classk is categorized as High -loaded).

63) When ($\hat{x}_{t+1}^i$ is classified as High) and ($\hat{y}_{t+1}^i$ is classified as Medium) and ($\hat{z}_{t+1}^i$ is classified as low) and ($\hat{w}_{t+1}^i$ is classified as High), then (classk is categorized as High-loaded).
64) When ($\hat{x}_{t+1}^i$ is classified as High) and ($\hat{y}_{t+1}^i$ is classified as Medium) and ($\hat{z}_{t+1}^i$ is classified as High) and ($\hat{w}_{t+1}^i$ is classified as low), then (classk is categorized as High-loaded).
65) When ($\hat{x}_{t+1}^i$ is classified as High) and ($\hat{y}_{t+1}^i$ is classified as High) and ($\hat{z}_{t+1}^i$ is classified as low) and ($\hat{w}_{t+1}^i$ is classified as Medium), then (classk is categorized as High-loaded).
66) When ($\hat{x}_{t+1}^i$ is classified as High) and ($\hat{y}_{t+1}^i$ is classified as High) and ($\hat{z}_{t+1}^i$ is classified as Medium) and ($\hat{w}_{t+1}^i$ is classified as low), then (classk is categorized as High-loaded).
67) When ($\hat{x}_{t+1}^i$ is classified as High) and ($\hat{y}_{t+1}^i$ is classified as Medium) and ($\hat{z}_{t+1}^i$ is classified as Medium) and ($\hat{w}_{t+1}^i$ is classified as High), then (classk is categorized as High-loaded).
68) When ($\hat{x}_{t+1}^i$ is classified as High) and ($\hat{y}_{t+1}^i$ is classified as High) and ($\hat{z}_{t+1}^i$ is classified as Medium) and ($\hat{w}_{t+1}^i$ is classified as Medium), then (classk is categorized as High-loaded).
69) When ($\hat{x}_{t+1}^i$ is classified as Medium) and ($\hat{y}_{t+1}^i$ is classified as High) and ($\hat{z}_{t+1}^i$ is classified as High) and ($\hat{w}_{t+1}^i$ is classified as Medium), then (classk is categorized as High-loaded).
70) When ($\hat{x}_{t+1}^i$ is classified as Medium) and ($\hat{y}_{t+1}^i$ is classified as Medium) and ($\hat{z}_{t+1}^i$ is classified as High) and ($\hat{w}_{t+1}^i$ is classified as High), then (classk is categorized as High-loaded).
71) When ($\hat{x}_{t+1}^i$ is classified as High) and ($\hat{y}_{t+1}^i$ is classified as Medium) and ($\hat{z}_{t+1}^i$ is classified as High) and ($\hat{w}_{t+1}^i$ is classified as Medium), then (classk is categorized as High-loaded).
72) When ($\hat{x}_{t+1}^i$ is classified as Medium) and ($\hat{y}_{t+1}^i$ is classified as High) and ($\hat{z}_{t+1}^i$ is classified as Medium) and ($\hat{w}_{t+1}^i$ is classified as High), then (classk is categorized as High-loaded).
73) When ($\hat{x}_{t+1}^i$ is classified as low) and ($\hat{y}_{t+1}^i$ is classified as High) and ($\hat{z}_{t+1}^i$ is classified as High) and ($\hat{w}_{t+1}^i$ is classified as High), then (classk is categorized as Over-loaded).
74) When ($\hat{x}_{t+1}^i$ is classified as High) and ($\hat{y}_{t+1}^i$ is classified as low) and ($\hat{z}_{t+1}^i$ is classified as High) and ($\hat{w}_{t+1}^i$ is classified as High), then (classk is categorized as Over-loaded).
75) When ($\hat{x}_{t+1}^i$ is classified as High) and ($\hat{y}_{t+1}^i$ is classified as High) and ($\hat{z}_{t+1}^i$ is classified as low) and ($\hat{w}_{t+1}^i$ is classified as High), then (classk is categorized as Over-loaded).
76) When ($\hat{x}_{t+1}^i$ is classified as High) and ($\hat{y}_{t+1}^i$ is classified as High) and ($\hat{z}_{t+1}^i$ is classified as High) and ($\hat{w}_{t+1}^i$ is classified as low), then (classk is categorized as Over-loaded).
77) When ($\hat{x}_{t+1}^i$ is classified as High) and ($\hat{y}_{t+1}^i$ is classified as High) and ($\hat{z}_{t+1}^i$ is classified as High) and ($\hat{w}_{t+1}^i$ is classified as Medium), then (classk is categorized as Over-loaded).
78) When ($\hat{x}_{t+1}^i$ is classified as High) and ($\hat{y}_{t+1}^i$ is classified as High) and ($\hat{z}_{t+1}^i$ is classified as Medium) and ($\hat{w}_{t+1}^i$ is classified as High), then (classk is categorized as Over-loaded).
79) When ($\hat{x}_{t+1}^i$ is classified as High) and ($\hat{y}_{t+1}^i$ is classified as Medium) and ($\hat{z}_{t+1}^i$ is classified as High) and ($\hat{w}_{t+1}^i$ is classified as High), then (classk is categorized as Over-loaded).
80) When ($\hat{x}_{t+1}^i$ is classified as Medium) and ($\hat{y}_{t+1}^i$ is classified as High) and ($\hat{z}_{t+1}^i$ is classified as High) and ($\hat{w}_{t+1}^i$ is classified as High), then (classk is categorized as Over-loaded).
81) When ($\hat{x}_{t+1}^i$ is classified as High) and ($\hat{y}_{t+1}^i$ is classified as High) and ($\hat{z}_{t+1}^i$ is classified as High) and ($\hat{w}_{t+1}^i$ is classified as High), then (classk is categorized as Over-loaded).

- Fuzzy Logic Application:

Once the server data is input into the system, fuzzy logic utilizes the defined rules and membership functions to generate fuzzy outputs for each data point, as illustrated in Figure 7. These outputs typically include the membership percentages of the server in each fuzzy category.

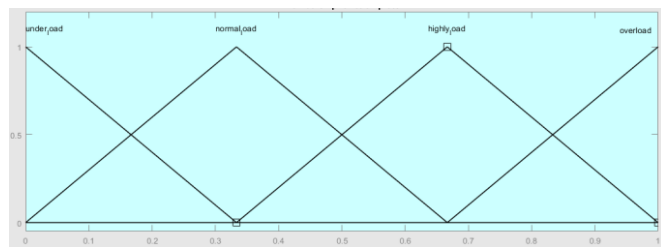

Figure. 7. Output Membership Function for Class Range [-0.6, 0.4]}

- Final Classification:

By combining various fuzzy outputs and employing post-processing logic, servers are categorized into different groups. The fuzzy system classifies servers into four levels: under-load, normal-load, highly-load, and over-load. Ultimately, the selection of the algorithm and settings for the fuzzy system in server classification depends on the environment and specific requirements of the system. In this fuzzy system, the Mamdani method has been utilized. This structured approach to classification ensures that server loads are accurately assessed, enabling effective load balancing and resource management within the network.

*Load Distribution Among Servers*

In the final stage of balancing server loads, as indicated in the algorithm 1, load distribution is performed among the servers present in the domain. Initially, as described in lines 1 to 3 of the algorithm, the consumption levels of CPU, memory, hard disk, and bandwidth for the servers are collected. Then, using the Long Short-Term Memory (LSTM) prediction algorithm, the estimated consumption of CPU, memory, hard disk, and bandwidth for each server is forecasted. The fuzzy system is then used to determine the load levels of the servers.

When a traffic flow enters the network, as shown in lines 8 to 13, servers classified as under-load are prioritized for receiving the traffic. Following that, servers at the normal-load level have a lower priority than under-load servers for traffic reception. Finally, servers at the highly-load level are selected as target servers, provided they do not fall into the over-load category.

Load distribution among servers requires transferring load from one server to another. If any server in the domain is at the over-load level, as described in lines 14 to 35, the load transfer will occur if the destination server meets one of the following conditions:

- The destination server is under-loaded and remains so after receiving the load. Servers meeting this condition have the highest priority for load reception.
- The destination server is under-load and, after the load transfer, transitions to a normal-load level. This condition, being the second highest priority, applies if no server meets the first condition.
- The destination server is at normal-load and remains at the same normal-load level after receiving the load. Finally, servers meeting this condition have the lowest priority.

If no servers meet these conditions and there are powered-off servers within the domain that can save energy, those servers will be powered on again, and the load from the over-load server will be transferred to the newly activated server. If other domains are available for load transfer, traffic will be redirected to another domain. If no powered-off servers are available and no servers in other domains can receive load, a new server will be added to the domain based on the controller's capacity.

If a server in the domain is at the highly-load level, it indicates that the server is on the verge of becoming overloaded. As described in the algorithm, load transfer from one server to another is performed similarly to when a server is in the over-load state. However, if no powered-off servers are available and no servers in other domains can receive load, the load transfer operation will not occur, and the domain will not be expanded.

If a server is under-load and another server in the domain is also under-load and can remain under-load after receiving the load, the load transfer will occur, and the source server will be powered off. This process also contributes to energy savings.

If a server is at normal-load, it indicates that the server is operating at its optimal state. The goal is to maintain all servers in the domain at this level through effective load distribution.

**Algorithm :**
Server Load Balancing:

**Input**: $C_j$, $Ld_{CPU}$, $Ld_{mem}$, $Ld_{hdd}$, $Ld_{bw}$, $h_i$
**Output**: D, T

1. **For** each server $h_i$ in the set of all servers
2. $$x_i = \frac{Ld_{CPU}}{\vartheta_i^x}, \quad 0 \leq x_i \leq 1$$
$$y_i = \frac{Ld_{mem}}{\vartheta_i^y}, \quad 0 \leq y_i \leq 1$$
$$z_i = \frac{Ld_{hdd}}{\vartheta_i^z}, \quad 0 \leq z_i \leq 1$$
$$w_i = \frac{Ld_{bw}}{\vartheta_i^w}, \quad 0 \leq w_i \leq 1$$
3. **End for**
4. Calculate $f_{CPU}$ using formula 5.
5. **Input** $f_{CPU}$ into the LSTM algorithm. Forecast output for all domain servers:
$\hat{x}_{t+1}^i$, $\hat{y}_{t+1}^i$, $\hat{z}_{t+1}^i$, $\widehat{w}_{t+1}^i$.
6. Classify servers using a fuzzy system based on predicted consumption:
$\hat{x}_{t+1}^i$, $\hat{y}_{t+1}^i$, $\hat{z}_{t+1}^i$, $\widehat{w}_{t+1}^i$.
7. Classification levels: under-load, normal-load, highly-load, over-load.
8. **For** each server $h_i$ in under-load, normal-load, or highly-load class
9.     Add traffic to the server and calculate the load after migration $\bar{h}_t$
10.     **If** $\bar{h}_t$ belongs to under-load, normal-load, or highly-load class *normal-load*.
11.         Direct new flow to server $h_i$.
12.     **End if**

| | |
|---|---|
| 13. | **End for** |
| 14. | **If** level of $h_i$ == over-load |
| 15. | **For** each server $h_i$ in the set of all servers $H$ |
| 16. | **If** level of $h_i$ == under-load or normal-load |
| 17. | $\tau \leftarrow h_i$ |
| 18. | **End if** |
| 19. | **End for** |
| 20. | **For** each server $h_i$ in $\tau$ |
| 21. | Calculate CPU, memory, hard disk, and bandwidth consumption based on formulas 1-4 and determine new levels. |
| 22. | **If** $h_i$== *under-load* and $\bar{h}_i$== *under-load* |
| 23. | T=$h_i$ |
| 24. | **End if** |
| 25. | **If** $h_i$== *under-load* and $\bar{h}_i$== *normal-load* |
| 26. | T=$h_i$ |
| 27. | **End if** |
| 28. | **If** $h_i$== *normal-load* and $\bar{h}_i$== *normal-load* |
| 29. | T= $h_i$ |
| 30. | **End if** |
| 31. | **Else** |
| 32. | Finally, turn on one of the domain servers. If none exist, add a new server. |
| 33. | **End if** |
| 34. | **End for** |
| 35. | **End if** |
| 36. | **If** level of $h_i$== highly-load |
| 37. | Repeat steps (14 to 35) for normal servers. If a server is off, turn it on; otherwise, do not transfer the load. |
| 38. | **End if** |
| 39. | **If** level of $h_i$== under-load |
| 40. | **For** each server $h_i$ in the set of all servers $H$ |
| 41. | **If** level $h_i$ == under-load or normal-load |
| 42. | $\tau \leftarrow h_i$ |
| 43. | **End if** |
| 44. | **End for** |
| 45. | **For** $h_i \in \tau$ |
| 46. | Calculation of CPU, memory, hard disk and bandwidth consumption according to formulas 1-4 for selected servers and determining the level of servers after load transfer. |
| 47. | **If** $h_i$== under-load and $\bar{h}_i$ == under-load |
| 48. | T=$h_i$ |
| 49. | **End if** |
| 50. | **End for** |
| 51. | **End if** |

For example, in the case of server load balancing, as illustrated in Figure 1, we have four servers: (server1, server2, server3, server4). Let's consider a scenario where the load levels of the servers (server1, server2, server3, server4) are (under-load, normal-load, highly-load, over-load ), respectively. In this situation, we first examine the server that is in the over-load state, as servers in the over-load state have higher priority for assessment. To transfer the load from this server to the others, servers two and three, which are in the under-load and normal-load states, respectively, are considered target servers. After transferring the load from the source server (server four) to the target servers (servers one and two), the load levels of these target servers are measured. The load levels of the target servers (servers one and two) after the transfer are both at normal-load.

According to the algorithm and the three conditions, since server one transitions from under-load to normal-load, it is selected as a target server. After the load transfer, the load levels of the servers in the domain become (normal-load, normal-load, highly-load, normal-load). In this state, since server three is at the highly-load level, it is designated as the source server for load transfer. Servers one, two, and three are then considered target servers. After the load transfer and determining the levels of these three servers, the load levels become (highly-load, normal-load, highly-load). As a result, only server two is considered the target server for the next transfer. After this transfer, the load status of the servers in the domain is equal to (normal-load, normal-load, normal-load, normal-load). This example illustrates the dynamic process of load balancing among servers, demonstrating how the load is redistributed to achieve optimal performance across the system.

## 4. Performance Evaluation

To simulate the proposed method, the Mininet emulator and the Floodlight controller were utilized. The OpenFlow protocol was employed for communication between the controller and the switches. Mininet [19] is a virtual testing platform that facilitates the development and testing of network tools and protocols. It effectively implements software-defined networks using Open vSwitch, allowing for the definition of multiple switches, each capable of running Open vSwitch.

Floodlight [20] is a widely used open-source software-defined networking (SDN) controller in research, education, and industry. Implemented in Java, it provides a modular and extensible platform for building software-defined network applications. The Floodlight controller uses the OpenFlow protocol for communication with switches and for managing packet forwarding in the data plane. It offers various features and capabilities for network traffic control, including flow-based transport, traffic engineering, monitoring, and security.

OpenFlow is a protocol standardized by the Open Networking Foundation (ONF) for software-defined networking. In this simulation, Floodlight communicates with switches using OpenFlow 1.3. sFlow is a network monitoring technology used to collect information from network devices. This technology is typically employed for monitoring network traffic, analyzing system performance, and identifying network issues. sFlow collects data directly from network equipment such as switches and routers.

sFlow-RT is a real-time analytics system that provides reliable information for executing DevOps, orchestration, and SDN operations. This system leverages sFlow technology to enable scalable monitoring of network infrastructures, computing resources, and applications using industry-standard sFlow tools.

For simulating multimedia flow in a software-defined Internet of Things (IoT) network, the Iperf tool was utilized. Iperf is a widely used open-source network performance testing tool that measures the maximum achievable bandwidth and throughput between two network endpoints. In Figure 8, a script depicting the network topology designed for load balancing is shown. This topology consists of four switches, one client, and four servers, with the primary goal of establishing load balancing among these four servers.

The simulation was conducted using Mininet, which is connected to the Floodlight controller. The Floodlight controller communicates with the switches via OpenFlow 1.3. Iperf was also employed to generate traffic in the network. In this network, the different servers have varying capacities. The load on the servers in this research was measured using sFlow. This comprehensive setup allows for a thorough evaluation of the proposed load-balancing method, ensuring its effectiveness in real-world scenarios within a controlled environment

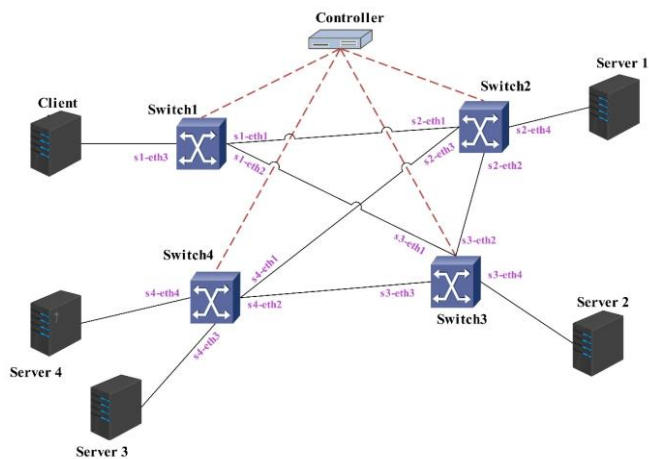

Figure. 8.    Network Topology for Performance Evaluation of the Server Load Balancing Algorithm

In this section, as illustrated in Figure 9, a scenario has been devised for simulating and evaluating server load balancing methods in software-defined Internet of Things (IoT) networks. This scenario consists of 26 streams, where the network streams begin at a volume of 10 megabytes and increase in increments of 10 megabytes until reaching a final stream size of 260 megabytes. By applying this scenario, the performance of various server load balancing methods can be tested under different conditions and scales. These experiments contribute to a more effective evaluation and comparison of different methods for achieving load balancing in servers and improving overall network performance.

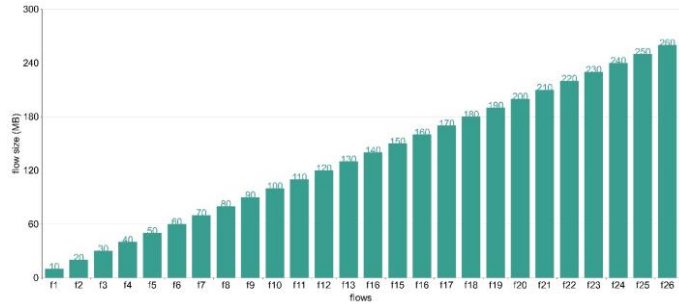

Figure. 9.    flow size Distribution in the Server Load Balancing Scenario.

In this section, to simulate multimedia flow in software-defined IoT networks, the Iperf tool was used to establish a UDP connection from the client to the other servers. Using this method, multimedia flows are sent from the server to other servers based on different algorithms. In a UDP connection, the size of the flow depends on the number of datagrams sent. Figure 10 displays the number of datagrams sent for each flow in the scenario described in the previous section

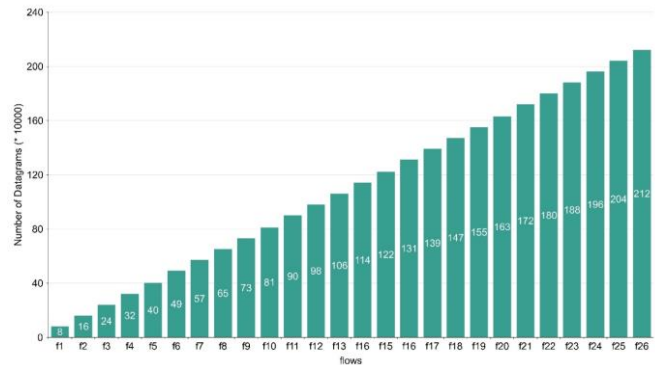

Figure. 10.    Number of Datagrams Sent per flow.

In this network, the servers have different capacities. Table 3 shows the capacity of each server.

Table 3.    Specifications and Capacities of the Servers Used in the Simulation

| *Servers* | *CPU* | *Memory* | *Hard Disk* | *Band Width* |
|---|---|---|---|---|
| *Server1* | 1GHz | 500MB | 900MB | 20MB |
| *Server2* | 2GHz | 550MB | 2GB | 20MB |
| *Server3* | 1GHz | 420MB | 800MB | 20MB |
| *Server4* | 1GHz | 450MB | 700MB | 20MB |

In the evaluation, traffic generation and server sizes were selected as key parameters to analyze the effectiveness of the proposed load-balancing approach. Traffic generation follows a progressive increase in flow sizes, simulating real-world network demand variations and enabling a systematic assessment of load distribution strategies under increasing stress conditions. Additionally, server sizes were defined to represent a heterogeneous computing environment, reflecting the diverse resource capacities commonly found in cloud and edge computing infrastructures. This heterogeneity ensures a realistic evaluation of how different load-balancing methods adapt to varying computational resources. These criteria provide a robust framework for assessing both the adaptability and efficiency of the proposed method in dynamic network conditions.

After executing the load balancing algorithms on the scenario, as shown in Figure 11, the servers that are at the under-load level and are not receiving any traffic are marked with an asterisk (*). The servers indicated with an arrow are designated as target servers for receiving traffic.

Initially, all four servers are in an under-load state, active, and operational. According to the load balancing algorithm, since there is more than one under-loaded server, two of the servers with the lowest load will be turned off. Consequently, two servers remain active, and the network manages with this reduced number of servers. As network traffic increases, the servers gradually reach the normal-load level. However, at the beginning of this scenario, none of the servers ever reach the over-load or highly-load levels, so there is no need to turn on the inactive servers. Thus, until the seventeenth stream, turning off two servers not only saves energy but also deactivates the four switches connected to those servers, contributing to energy efficiency in the network.

In flow 18, the second server reaches the highly-load level, and based on previous predictions, the third server is turned on to prepare for load transfer from the overloaded server. With the transfer of flow 19 to the first server, the excess load from the second server is shifted to the third server, achieving balance, while the first server also becomes overloaded. flow 20 is directed to the third server, which also reaches the highly-load level upon accepting the traffic, and since this situation was anticipated, the fourth server is turned on. Concurrently, traffic is transferred from the first server to the third, resulting in the first server returning to normal-load, while the second server remains at the same level.

Flow 21 moves towards the second server, causing it to reach the highly-load level. Simultaneously, as new traffic is sent to the second server, excess traffic from the third server is transferred to the fourth server, allowing the third server, which was at the highly-load level, to return to normal-load, while the fourth server, which was under-load, reaches normal-load. flow 22 is directed towards the fourth server, which then reaches the highly-load level after receiving the traffic. At the same time, excess traffic from the second server is initially transferred to the third server; however, despite this transfer, the second server remains at the highly-load level. Therefore, a second transfer from the second server to the first takes place, ultimately bringing the second server down to normal-load.

Flow 23 heads towards the second server, causing it to reach the highly-load level after accepting this stream. Concurrently, excess traffic from the fourth server is initially transferred to the third server; however, the fourth server remains at the highly-load level. Thus, a second transfer from the fourth server to the second server occurs, and eventually, the fourth server returns to normal-load. With the arrival of flow 24 at the third server, this server reaches the highly-load level. Concurrently, since the first server has reached the highly-load level after accepting flow 23 and requires load transfer, but there are no target servers available to accept the excess load (as servers two and four would also reach the highly-load level with this traffic), the first server remains at the highly-load level. Upon the arrival of flow 25 at the second server, this server also reaches the highly-load level. At the same time, load balancing for the third server is not feasible, similar to the first server. Furthermore, with flow 26 directed to the fourth server, this server also reaches the highly-load level. At this point, all four servers are at the highly-load level, yet none of them have reached the over-load level.

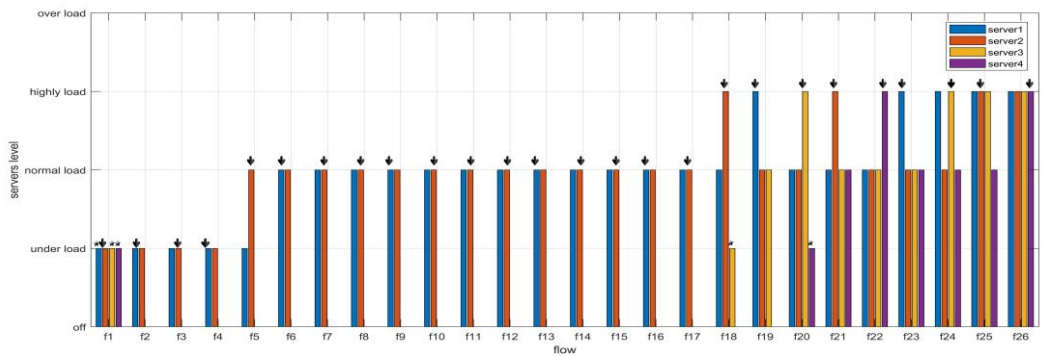

Figure. 11.   Server Levels in the Load Balancing Scenario

At this stage, the proposed scheme is compared with two other methods for server selection: random selection and round-robin selection. These methods were chosen due to their widespread use as baseline techniques in load balancing. Random selection provides a simple yet uncontrolled distribution of traffic, while round-robin ensures fairness but does not consider real-time server load. Evaluating our proposed method against these well-established approaches highlights its advantages in dynamic load prediction and energy efficiency. Additionally, these methods are commonly used in practical SDN and cloud environments, ensuring a relevant and meaningful comparison. Figure 12 illustrates the server levels after executing the random selection scheme. As seen in this figure, after server four receives flow 12, it reaches the over-load level but is still selected as the target server for accepting flows. Ultimately, with the arrival of flow 26, three servers reach the over-load level, while the third server remains at the normal-load level.

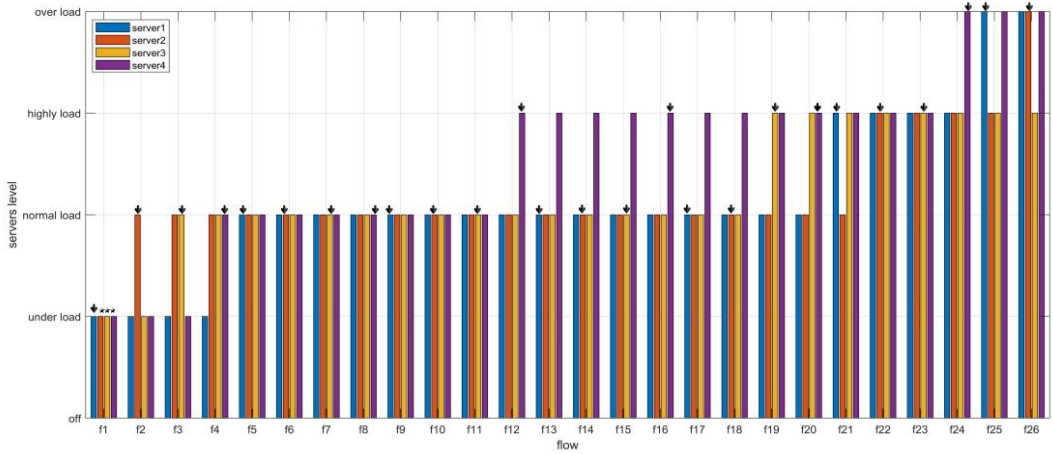

Figure. 12.    Server Load Levels in the Round-Robin Selection Scheme

Figure 13 also shows the server levels after executing the round-robin selection scheme. Server four reaches the over-load level after processing flow 13; however, despite this status, three other streams are sent to this server. In the final stream, which is flow 26, both the third and fourth servers are at the over-load level, while the first and second servers are at the normal-load level.

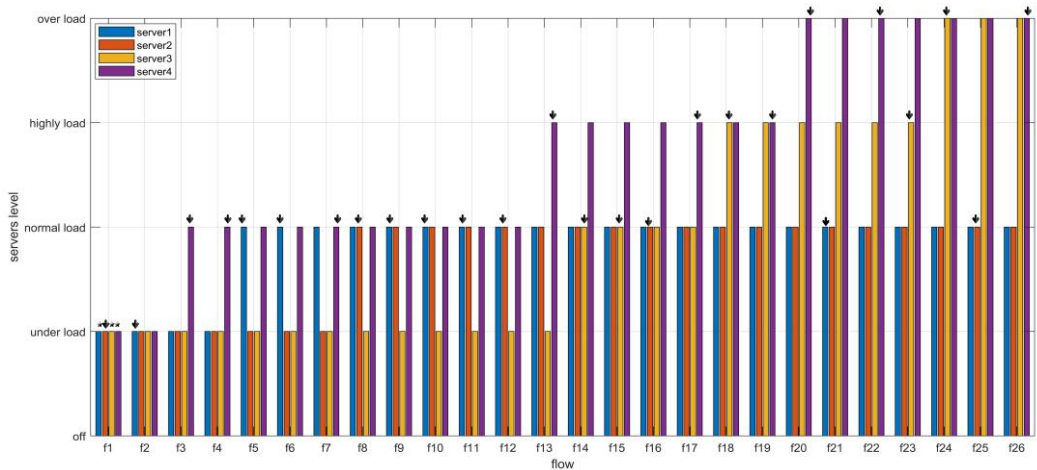

Figure. 13.    Server Load Levels in the Round-Robin Selection Scheme

In the proposed framework, the Long Short-Term Memory (LSTM) algorithm is utilized for prediction, as this algorithm is suitable for forecasting time series with unknown lag sizes. Moreover, LSTM allows us to access information from multiple previous time steps instead of just the immediate past, thanks to its use of recurrent deep networks. The number of steps to look back can be adjusted based on the type of input data.

In the proposed approach, predictions are made for the CPU, memory, hard disk, and bandwidth utilization of each server. For each metric being predicted, the number of steps to look back varies. For predicting CPU usage, hard disk usage, and bandwidth, a look-back of 5 is considered, while for memory prediction, a look-back of 2 is applied. The look-back values were fine-tuned using a trial-and-error method to determine the most suitable settings.

In Figure 14, you can observe the predicted CPU usage of the second server using the LSTM prediction algorithm. In this figure, the blue lines represent actual CPU usage, while the orange and green lines indicate the predicted values.

For the predictions made using the LSTM algorithm, three metrics are employed to demonstrate the accuracy of the predictions. The first metric is the Root Mean Square Error (RMSE); a lower RMSE value indicates better predictions. The second metric is the Mean Absolute Error (MAE), which reflects the absolute error between the predicted and actual values. The third metric is the coefficient of determination ($R^2$), a statistical measure that indicates the percentage of variance or dispersion in the data. The $R^2$ value ranges between zero and one; an $R^2$ of one means that the predictions perfectly match the actual data, while an $R^2$ of zero indicates no correlation, akin to a horizontal line. Values between zero and one suggest varying degrees of model performance, with higher values indicating better performance.

Additionally, the LSTM model is implemented using the Sequential API from TensorFlow/Keras, consisting of one LSTM layer with 5 neurons and a Dense output layer with one neuron. The lookback window for input data varies based on the resource metric, with values empirically determined to optimize performance. Training is performed using the Adam optimizer with a Mean Squared Error (MSE) loss function, a batch size of 1, and 15 epochs. MinMaxScaler is applied for normalization, and the dataset is split into 80% training and 20% testing. This setup ensures effective time-series forecasting for server load prediction in software-defined networks.

The dataset used for training and testing the LSTM model was collected from real-world server logs in a software-defined networking (SDN) environment. It includes system resource utilization metrics such as CPU usage, memory usage, hard disk activity, and bandwidth consumption. The dataset consists of approximately $N$ records spanning a time period of $T$ hours/days, with each record containing time-series observations of multiple resource utilization metrics. Pre-processing steps included handling missing values using linear interpolation and applying MinMax scaling to normalize the data within the range [0,1], ensuring stable model training. No artificial data augmentation techniques were applied, as the dataset already provides sufficient variability in load conditions.

The LSTM model consists of two hidden layers, each with 64 hidden units. A batch size of 32 was used during training. The learning rate was set to 0.001 and optimized using a decay schedule. The Adam optimizer was used for training due to its efficiency in handling time-series data. The model was trained using Mean Squared Error (MSE) as the loss function. These hyperparameters were fine-tuned through multiple experiments to achieve optimal prediction accuracy while avoiding overfitting.

In predicting the CPU usage of the second server using the LSTM algorithm, the RMSE was found to be 7.75, the MAE was 6.31, and the R² value was 0.77. These results indicate that the LSTM algorithm performs very well in predicting CPU usage.

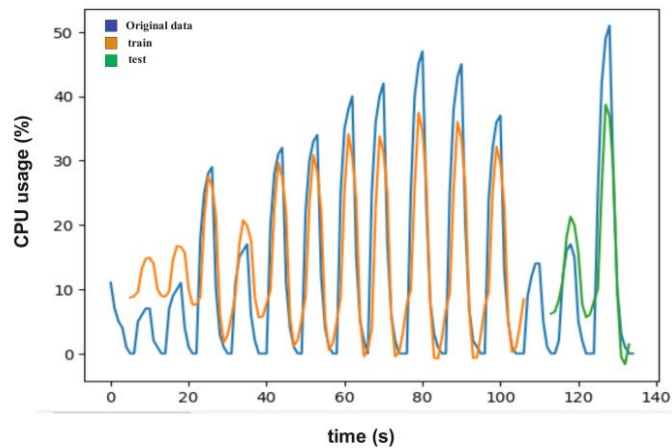

Figure. 14. Predicted CPU Usage of the Second Server Using the LSTM Algorithm

In Figure 15, you can see the predicted memory usage of the second server using the LSTM algorithm. In this figure, the blue lines represent actual memory usage, while the orange and green lines indicate the predicted amounts. For the memory usage prediction of the second server, the RMSE was 19, the MAE was 14.67, and the $R^2$ value was 0.51. These results show that the LSTM algorithm has good performance for predicting memory usage.

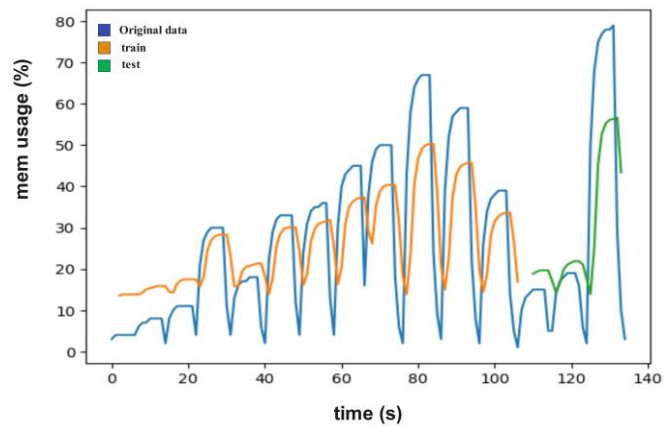

Figure. 15. Predicted Memory Usage of the Second Server Using the LSTM Algorithm

In Figure 16, the predicted hard disk usage of the second server using the LSTM algorithm is shown. The blue lines represent actual hard disk usage, while the orange and green lines represent the predicted values. In this prediction, the RMSE was 5.50, the MAE was 3.84, and the $R^2$ value was 0.6. These results indicate that the LSTM algorithm performs excellently in predicting hard disk usage.

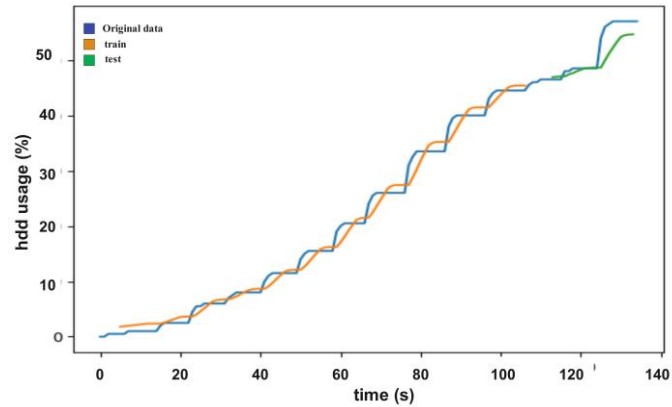

Figure. 16.    Predicted Hard Disk Usage of the Second Server Using the LSTM Algorithm

In Figure 17, you can observe the predicted bandwidth usage of the second server using the LSTM algorithm. In this figure, the blue lines represent actual bandwidth usage, while the orange and green lines show the predicted values. In predicting the bandwidth usage of the second server, the RMSE was 22.05, the MAE was 15.79, and the $R^2$ value was 0.67. These results indicate that the LSTM algorithm performs very well in predicting bandwidth usage.

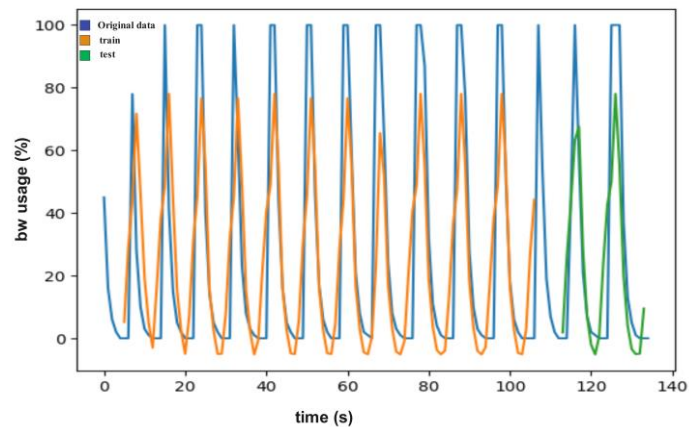

Figure. 17.    Predicted Bandwidth Usage of the Second Server Using the LSTM Algorithm

## 5. Conclusions

The focus of this paper was to establish effective load distribution among servers within software-driven multimedia Internet-of-Things (IoT) networks. The suggested method successfully mitigated the risk of server overload by utilizing the Long Short-Term Memory (LSTM) algorithm for forecasting server usage. Following this, fuzzy logic was employed to assess the load on each server and allocate tasks accordingly. This approach led to enhanced load distribution throughout the network, contributing to improved energy efficiency.

**Declarations**

*Funding*
This research did not receive any grant from funding agencies in the public, commercial, or non-profit sectors.

*Authors' contributions*
SI: Study design, acquisition of data, interpretation of the results, statistical analysis, drafting the manuscript.

AM: Study design, interpretation of the results, drafting the manuscript, revision of the manuscript.

SA: Guidance on artificial intelligenc methods, interpretation of AI-related results.

## Conflict of interest

There are no conflicts of interest associated with this publication.